\documentclass[aps,pra,twocolumn,showpacs,amsmath,amssymb,longbibliography,superscriptaddress]{revtex4-1}

\usepackage[english]{babel}
\usepackage{amsmath,amsfonts,bm,graphicx,verbatim,mathrsfs,units,color}
\usepackage[colorlinks=true,citecolor=blue]{hyperref}
\usepackage{pdfpages}

\begin{document}

\title{Distribution of current fluctuations in a bistable conductor}
\author{S. Singh}
\affiliation{Low Temperature Laboratory, Department of Applied Physics,
Aalto University, 00076 Aalto, Finland}
\author{J. T. Peltonen}
\affiliation{Low Temperature Laboratory, Department of Applied Physics,
Aalto University, 00076 Aalto, Finland}
\author{I. M. Khaymovich}
\affiliation{Low Temperature Laboratory, Department of Applied Physics,
Aalto University, 00076 Aalto, Finland}
\affiliation{Institute for Physics of Microstructures, Russian Academy of Sciences, 603950 Nizhny Novgorod, GSP-105, Russia}
\affiliation{LPMMC, Universit\'{e} de Grenoble-Alpes and CNRS, 38042 Grenoble, France}
\author{J. V. Koski}
\affiliation{Low Temperature Laboratory, Department of Applied Physics,
Aalto University, 00076 Aalto, Finland}
\affiliation{Solid State Physics Laboratory, ETH Zurich, 8093 Zurich, Switzerland}
\author{C. Flindt}
\affiliation{Low Temperature Laboratory, Department of Applied Physics,
Aalto University, 00076 Aalto, Finland}
\author{J. P. Pekola}
\affiliation{Low Temperature Laboratory, Department of Applied Physics,
Aalto University, 00076 Aalto, Finland}

\date{\today}


\begin{abstract}
We measure the full distribution of current fluctuations in a single-electron transistor with a controllable bistability. The conductance switches randomly between two levels due to the tunneling of single electrons in a separate single-electron box. The electrical fluctuations are detected over a wide range of time scales and excellent agreement with theoretical predictions is found. For long integration times, the distribution of the time-averaged current obeys the large-deviation principle.  We formulate and verify a fluctuation relation for the bistable region of the current distribution.
\end{abstract}

\maketitle

\emph{Introduction.---} Nano-scale electronic conductors operated at low temperatures are versatile tools to test predictions from statistical mechanics \cite{Maruyama2009,Parrondo2015,Pekola2015}. The ability to detect single electrons in Coulomb-blockaded islands has in recent years paved the way for solid-state realizations of Maxwell's demon \cite{Koski2015,Chida2015}, Szilard's engine \cite{Koski2014}, and the Landauer principle of information erasure, complementing experiments with colloidal particles \cite{Toyabe2010,Berut2012,Jun2014}. Moreover, several fluctuation theorems have been experimentally verified in electronic systems, including observations of negative entropy production at finite times \cite{Nakamura2010,Utsumi2010,Kung2012,Saira2012,Koski2013}. Bistable systems constitute another class of interesting phenomena in statistical physics \cite{vanKampen2007}. Bistabilities can be found in many fields of science~\cite{Dieterich2015,Santo2016} and can for example be caused by external fluctuators or intrinsic non-linearities. Bistabilities lead to flicker noise which can be detrimental to the controlled operation of solid-state qubits \cite{Paladino2014} and other nano-devices whose fluctuations we wish to minimize~\cite{Gustafsson2013,Pourkabirian2014}.

In this Rapid Communication, we realize a controllable bistability that causes current fluctuations in a nearby conductor. The tunneling of electrons in a single-electron box (SEB) makes the conductance switch between two levels in a nearby single-electron transistor~(SET) whose current is monitored in real-time. With this setup, we can accurately measure the distribution of current fluctuations in a bistable conductor, including the exponentially rare fluctuations in the tails, and we can test fundamental concepts from statistical physics as we modulate the bistability in a controlled manner.

\begin{figure}[t]
\includegraphics[width=0.9\columnwidth]{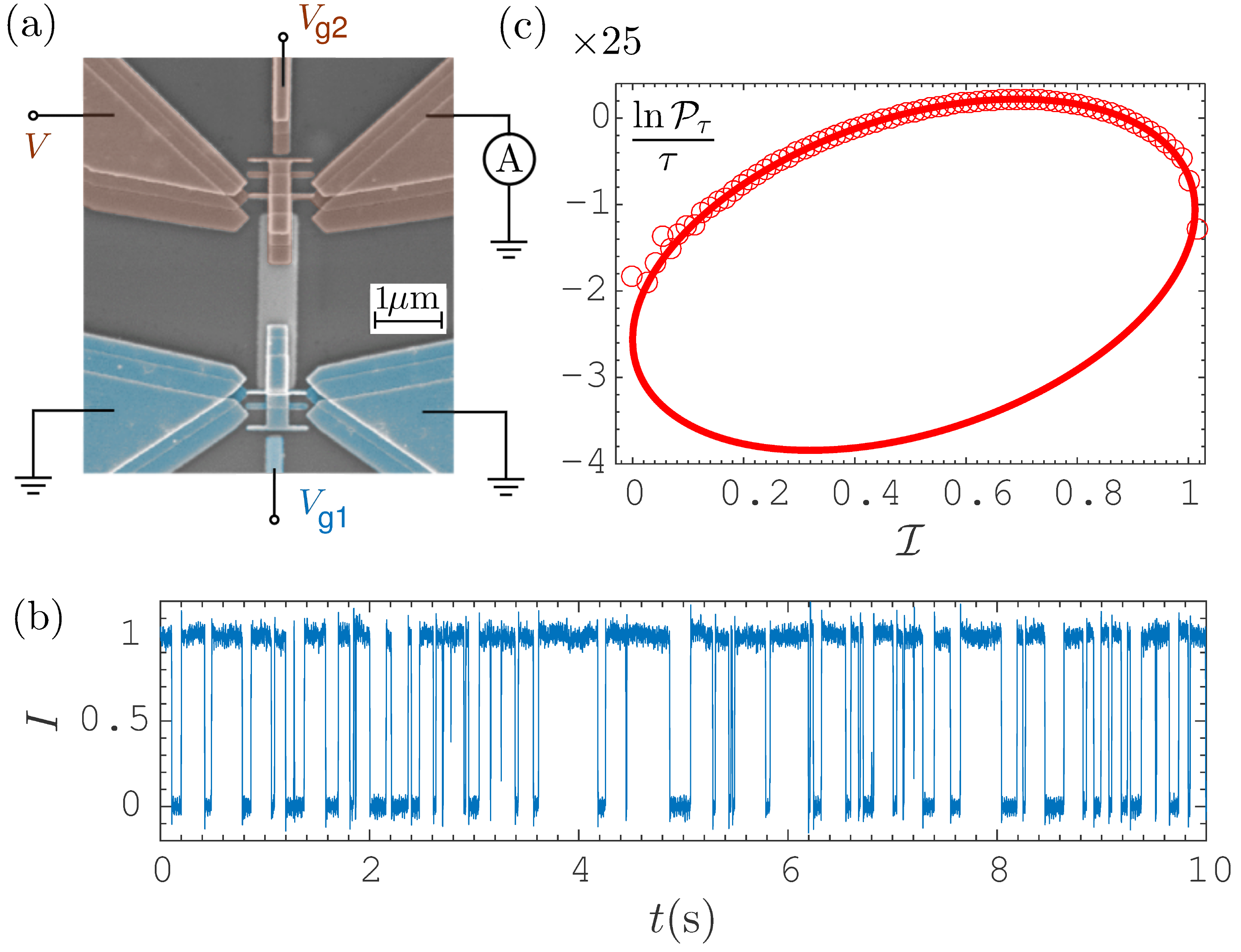}
\caption{(color online). Experimental setup. (a)~False-colored scanning electron micrograph of the SET (brown) and the SEB (blue). Both devices are fabricated by electron beam lithography and three-angle shadow evaporation~\cite{Joonas2015}. The gate voltages $V_\text{g1}$ and $V_\text{g2}$ are used to control the tunneling rates. The bias voltage $V$ is applied across the SET and the current $I$ is measured. (b)~The current in the SET switches between the two normalized values $\langle I_-\rangle~=0$ and $\langle I_+\rangle= 1$ due to the tunneling of single electrons on and off the SEB. (c)~Distribution of the time-integrated current (\textcolor{red}{$\circ$}) for the integration time $\tau=180$~ms together with the tilted ellipse given by Eq.~(\ref{eq:P(I)}). The controllable rates for tunneling on and off the SEB are $\Gamma_+= 72$ Hz and $\Gamma_-= 37 $ Hz.}
\label{fig:1}
\end{figure}

\begin{figure*}[t]
\includegraphics[width=0.9\textwidth]{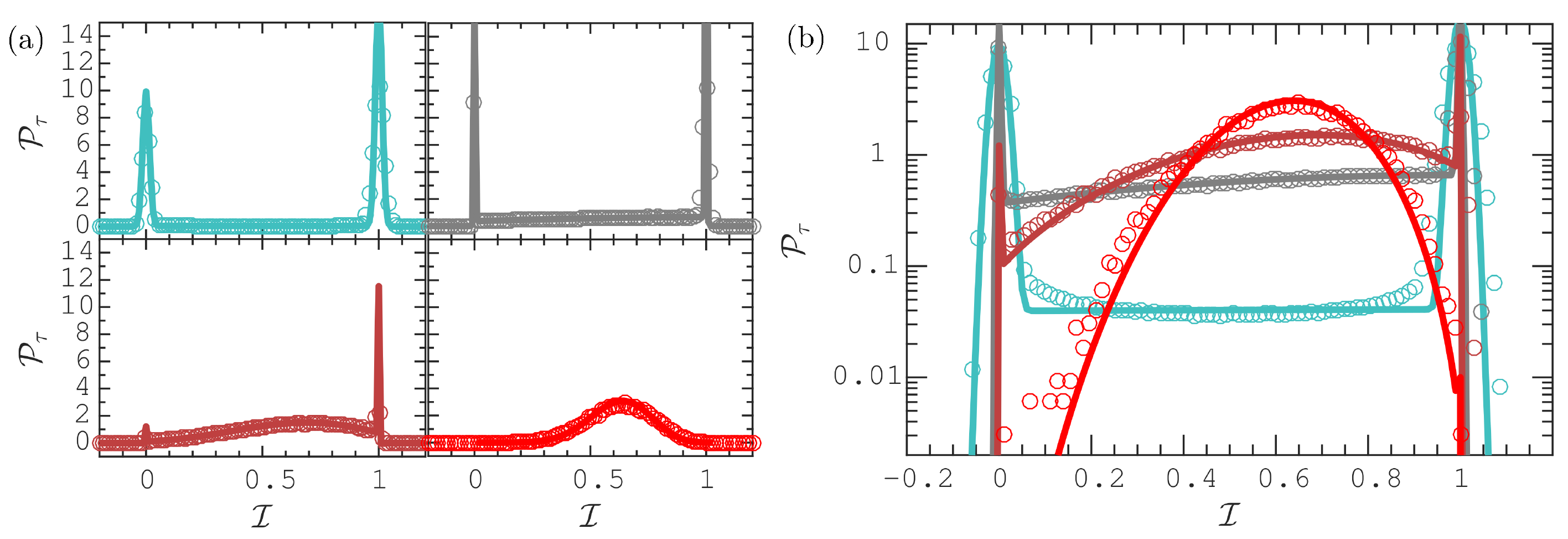}
\caption{(color online). Distribution of the time-integrated current. (a) Experimental (circles) and theoretical (lines) results for the distribution $\mathcal{P}_\tau(\mathcal{I})$ with the integration times, $\tau=$ 4 ms (left top panel), 80 ms (right top panel), 320 ms (left bottom panel), and 1280 ms (right bottom panel). The tunneling rates are $\Gamma_+=130$ Hz and $\Gamma_-=70$ Hz. The theory curves are based on Eqs.~(\ref{eq:I_tau},\ref{eq:P(N)}) with no adjustable parameters. For short times, the distribution is bimodal with distinct peaks around the normalized currents $\langle I_-\rangle~=0$ and  $\langle I_+\rangle= 1$. At long times, the distribution is approximately normal-distributed around the mean $\langle I\rangle=\Gamma_+/(\Gamma_++\Gamma_-)~\simeq 0.65$. (b) The same results on a logarithmic scale.}
\label{fig:2}
\end{figure*}

\emph{Experimental setup.---} Figure~\ref{fig:1}a shows an SET which is capacitively coupled to an SEB. Both are composed of small normal-conducting islands coupled to superconducting leads via insulating tunneling barriers. Measurements are performed at around 0.1~K, well below the charging energy of both the SEB and the SET. The tunneling rates of the SEB are tuned to the kilo-hertz regime so that the tunneling of electrons on and off the SEB are separated by milliseconds. The tunneling rates in the SET are on the order of several hundred megahertz and the electrical current is in the range of picoamperes. The conductance of the SET is highly sensitive to the presence of individual electrons on the SEB. This can be used to detect the individual tunneling events in the SEB by monitoring the current in the SET, see Fig.~\ref{fig:1}b \cite{Lu2003,Fujisawa2006,Gustavsson2006,Sukhorukov2007,Fricke2007,Gustavsson2009,Flindt2009,Ubbelohde2012,House2013,Maisi2014}.
Here, by contrast, we turn around these ideas and instead we focus on the current fluctuations in the SET under the influence of the random tunneling events in the SEB~\cite{Hassler2011}. This concept has an immediate application in characterization of spurious two-level fluctuators which appear in many solid-state devices \cite{Eiles1993, Paladino2014} and may affect the device properties. Thus, we use the SEB as an exemplary two-level fluctuator which can be completely characterized by considering the statistics of the current in the device in focus (the SET in our case).

\emph{Time-averaged current.---}  Our dynamical observable is the time-averaged current
\begin{equation}
\mathcal{I}(\tau)=\frac{1}{\tau}\int_{t_0}^{t_0+\tau}dt I(t)
\end{equation}
measured over the time interval $[t_0,t_0+\tau]$. For stationary processes, the distribution of current fluctuations depends only on the length of the interval $\tau$ and not on $t_0$. The distribution is expected to exhibit general properties that should be observable in any bistable conductor. For example, it has been predicted~\cite{Jordan2004} and verified in a related experiment~\cite{Sukhorukov2007} that the logarithm of the distribution at long times is always given by a tilted ellipse, see Fig.~\ref{fig:1}c and Refs.~\cite{Utsumi2007,Lambert2015}.
Besides, the crossover from short to long times, see Fig.~\ref{fig:2}, gives additional information about the bistable system (SEB). This information can be used for the detection and characterization of parasitic two-level fluctuators that may be present in the vicinity of a device, since the fluctuation statistics should be universal for all such two-level systems as we will see.

\emph{Measurements.---} Figure~\ref{fig:2}a shows experimental results for the distribution of the time-averaged current.  For short integration times, the distribution is bimodal with two distinct peaks centered on the average currents $\langle I_-\rangle$ and  $\langle I_+\rangle$ corresponding to having either $n=0$ or $n=1$ electrons on the SEB. According to the central limit theorem, the fluctuations should become normal-distributed with increasing integration time $\tau$, having a variance that decreases as $\mathcal{O}(\tau^{-1/2})$. This expectation is confirmed by Fig.~\ref{fig:2}a, showing how the distribution becomes increasingly peaked around the mean current~$\langle I\rangle$. This behavior is similar to the suppression of energy or particle fluctuations in the ensemble theories of statistical mechanics~\cite{Pathria2011}, here with the limit of long integration times playing the role of the thermodynamic limit. However, even as the distribution becomes increasingly peaked, rare fluctuations persist~\cite{Touchette2009}. This can be visualized by using a logarithmic scale which emphasises the rare fluctuations encoded in the tails of the distribution, Fig.~\ref{fig:2}b. The rare fluctuations will be important when we below formulate and test a fluctuation theorem.

\begin{figure*}[t]
\includegraphics[width=0.9\textwidth]{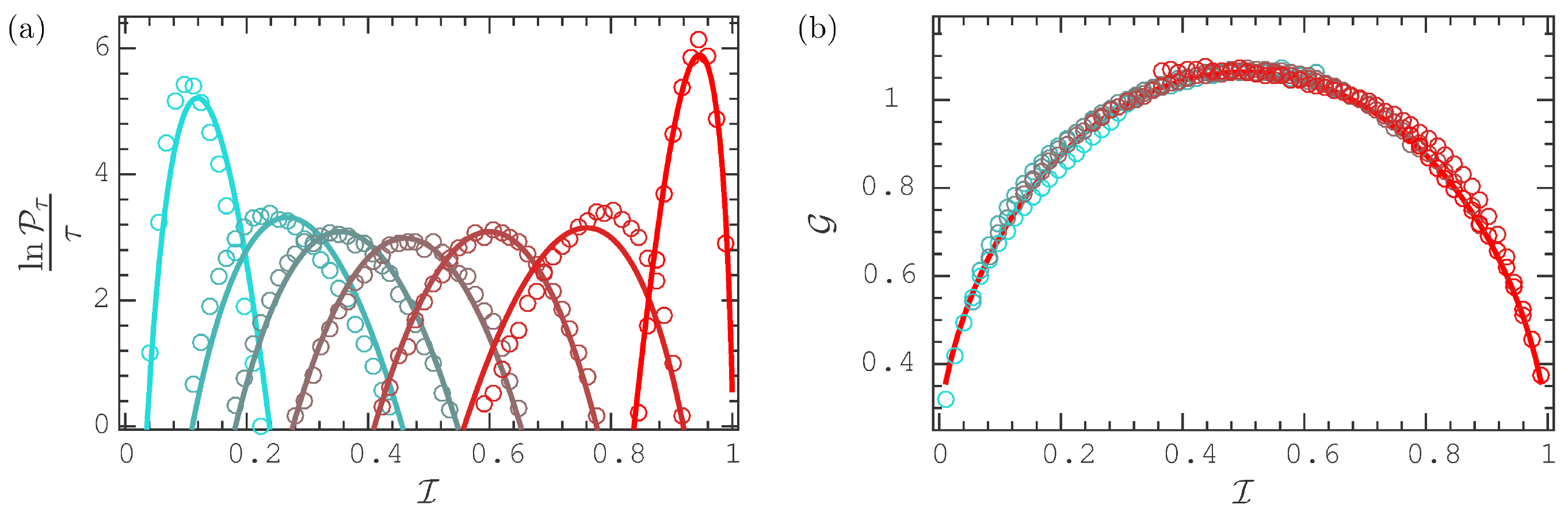}
\caption{(color online). Tilted ellipse and and universal semi-circle. (a)  Experimental (circles) and theoretical (lines) results for the logarithm of the distribution over the integration time, $\ln[\mathcal{P}_\tau(\mathcal{I})]/\tau$. The integration time is $\tau=400$ ms and the controllable tunneling rates $\Gamma_+ /\Gamma_- $ are 40/315 Hz (turquoise), 109/248, 85/187, 88/98, 120/81, 166/55 and 285/20  (red) from left to right. The curves obtained from Eq.~(\ref{eq:P(I)}) include a finite-time off-set which is inversely proportional to $\tau$~\cite{sup_mat}. (b) When rescaled according to Eq.~(\ref{eq:G(I)}), the experimental data collapse onto the universal semi-circle indicated with a full red line.}
\label{fig:3}
\end{figure*}

\emph{Theory.---} To better understand the fluctuations we develop a detailed theory of the distribution function $\mathcal{P}_\tau(\mathcal{I})$. The current fluctuates due to the random tunneling in the SEB and because of intrinsic noise in the SET itself. This we describe by the stochastic equation
\begin{equation}
\mathcal{I}(\tau)= [1-\mathcal{N}(\tau)]\langle I_-\rangle+\mathcal{N}(\tau)\langle I_+\rangle+\xi(\tau),
\label{eq:I_tau}
\end{equation}
where the first two terms account for the random switching between the average currents $\langle I_-\rangle$ and  $\langle I_+\rangle$ and we have defined the time-averaged electron number $\mathcal{N}(\tau)=\int_{t_0}^{t_0+\tau}dt n(t)/\tau$ on the SEB with $n=0,1$ \cite{Utsumi2007}. The time-averaged noise $\xi(\tau)$ describes the intrinsic fluctuations around the mean values $\langle I_-\rangle$ and $\langle I_+\rangle$, assumed here to be independent of $n$.  The distribution of current fluctuations $\mathcal{P}_\tau(\mathcal{I})$ is now determined by the fluctuations of $\mathcal{N}(\tau)$ and the intrinsic noise $\xi(\tau)$. Electrons tunnel on and off the SEB with the tunneling rates $\Gamma_+$ and $\Gamma_-$, changing $n$ from 0 to 1 and vice versa, respectively. The distribution  $\mathcal{P}_\tau(\mathcal{N})$ then becomes~\cite{sup_mat}

\begin{widetext}
\begin{equation}
\mathcal{P}_\tau(\mathcal{N}) = \frac{e^{-[\Gamma_+(1-\mathcal{N})+\Gamma_- \mathcal{N}]\tau }}{\Gamma_++\Gamma_-}\left\{\Gamma_-\delta\left(\mathcal{N}\right)+ \Gamma_+\delta\left(1-\mathcal{N}\right)+2\tau\Gamma_+\Gamma_-\Pi(\mathcal{N})\left[I_0(x) + \tau[\Gamma_-(1-\mathcal{N})+\Gamma_+ \mathcal{N}]\frac{I_1(x)}{x}\right]\right\}.
\label{eq:P(N)}
\end{equation}
\end{widetext}
The boxcar function $\Pi(\mathcal{N})= \theta(\mathcal{N})\theta(1-\mathcal{N})$ is given by Heaviside step functions and $I_{0,1}(x)$ are modified Bessel functions of the first kind. We have also introduced the dimensionless parameter $x=2\tau\sqrt{\Gamma_+\Gamma_-\mathcal{N}(1-\mathcal{N})}$
controlling the distribution profile. For $\Gamma_+=\Gamma_{-}$, we recover the result of Ref.~\cite{Bergli2009}. Based on $\mathcal{P}_\tau(\mathcal{N})$, we can evaluate $\mathcal{P}_\tau(\mathcal{I})$ according to Eq.~(\ref{eq:I_tau}), taking into account the intrinsic fluctuations given by $\xi(\tau)$.

The resulting theory curves agree well with the experimental data in Fig.~\ref{fig:2} over a wide range of integration times. For short times, $\tau\ll (\Gamma_++\Gamma_-)^{-1}$, we have $\mathcal{P}_\tau(\mathcal{N}) \simeq [\Gamma_-\delta\left(\mathcal{N}\right)+ \Gamma_+\delta\left(1-\mathcal{N}\right)]/(\Gamma_++\Gamma_-)$, corresponding to $\mathcal{P}_\tau(\mathcal{I})$ being bimodal with distinct peaks centered around the two average currents $\langle I_-\rangle$ and  $\langle I_+\rangle$. With increasing integration time, the distribution eventually takes on the large-deviation form $\mathcal{P}_\tau(\mathcal{N})\propto e^{G(\mathcal{N})\tau}$ with the  rate function $G(\mathcal{N})=-(\sqrt{\Gamma_+(1-\mathcal{N})}-\sqrt{\Gamma_-\mathcal{N}})^2$ following directly from Eq.~(\ref{eq:P(N)}). Moreover, the long-time limit of the distribution describes the low-frequency current fluctuations which should be dominated by the slow switching process. We can then ignore $\xi(\tau)$ in Eq.~(\ref{eq:I_tau}) such that the distribution becomes~\cite{Jordan2004}
\begin{equation}
\frac{\ln\mathcal{P}_\tau(\mathcal{I})}{\tau} \simeq -\frac{\left[\sqrt{\Gamma_+(\langle I_+\rangle-\mathcal{I})}-\sqrt{\Gamma_-(\mathcal{I}-\langle I_-\rangle)}\right]^2}{\langle I_+\rangle-\langle I_-\rangle}.
\label{eq:P(I)}
\end{equation}
The rate function on the right-hand side characterizes the non-gaussian fluctuations of the current beyond what is described by the central limit theorem \cite{Touchette2009}. Importantly, the rate function is independent of the integration time and it captures the exponential decay of the probabilities to observe rare fluctuations. Geometrically, the rate function describes the upper part of a tilted ellipse, delimited by the currents $\langle I_+\rangle$ and $\langle I_-\rangle$ \cite{Jordan2004,Sukhorukov2007,Utsumi2007,Lambert2015}. The tilt is given by the ratio of the controllable tunneling rates $\Gamma_+$ and $\Gamma_-$ and its width is governed by their product (see the prefactor in the parameter $x$). The tilted ellipse agrees well with the experimental results in the bistable range, $\langle I_-\rangle\leq \mathcal{I} \leq \langle I_+\rangle$, as seen in Fig.~\ref{fig:1}c. The extreme tails of the distribution are determined by the intrinsic fluctuations around $\langle I_-\rangle$ and $\langle I_+\rangle$ which are not included in Eq.~(\ref{eq:P(I)}).

By adjusting the tunneling rates we can control the shape and the tilt of the ellipse as illustrated in Fig.~\ref{fig:3}a. For $\Gamma_-\gg \Gamma_+$, the ellipse is strongly tilted to one side and the distribution is mostly centered around $\langle I_-\rangle$. As the ratio of the rates is changed, the ellipse becomes tilted to the other side and the distribution gets centered around $\langle I_+\rangle$. The average $\langle \mathcal{I} \rangle$ is given by the value of $\mathcal{I}$ where the distribution is maximal. This value changes from $\langle I_-\rangle$ to $\langle I_+\rangle$ as we tilt the ellipse. The abruptness of the change is determined by the width of the ellipse.

\begin{figure}
\includegraphics[width=0.9\columnwidth]{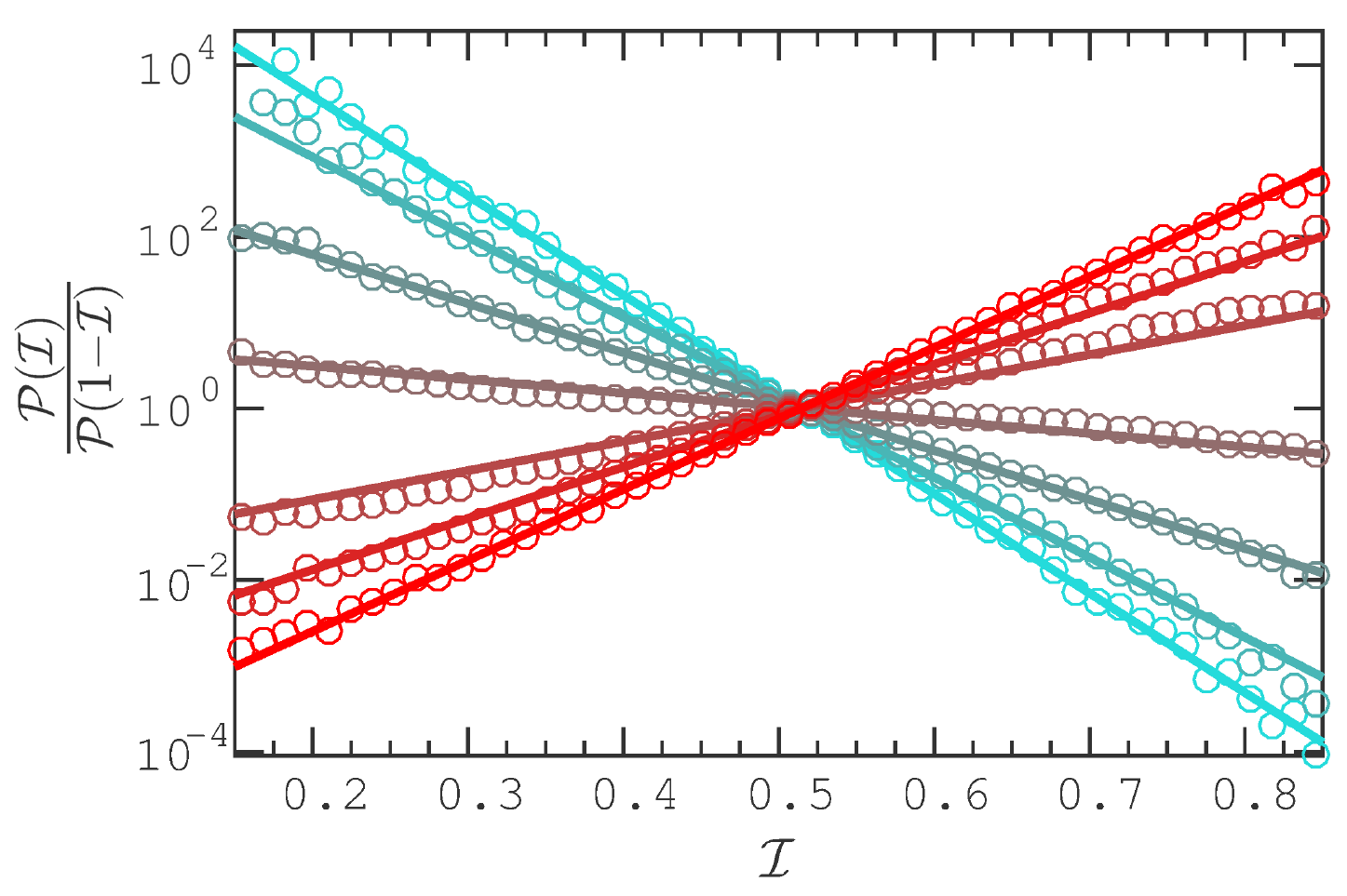}
\caption{(color online). Experimental test of the fluctuation relation. With the two normalized average currents $\langle I_+\rangle=1$ and $\langle I_-\rangle=0$, Eq.~(\ref{eq:FT}) can be brought on the equivalent form $P_\tau(\mathcal{I})/P_\tau(1-\mathcal{I})= \exp[2\tau(\Gamma_+-\Gamma_-)(\mathcal{I}-1/2)]$. In the figure, we plot the left-hand side using the experimental data ($\circ$) from Fig.~\ref{fig:3}, while the right-hand side is shown as full lines.  }
\label{fig:4}
\end{figure}
\emph{Universal semi-circle.---} To provide a unified description of the fluctuations we define the rescaled distribution
\begin{equation}
\mathcal{G}(\mathcal{I})\equiv\frac{1}{2\sqrt{\Gamma_+\Gamma_-}}\left\{\frac{\ln\mathcal{P}_\tau(\mathcal{I})}{\tau} + \frac{\Gamma_+\Delta\mathcal{I}_+-\Gamma_-\Delta\mathcal{I}_-}{\langle I_+\rangle-\langle I_-\rangle}\right\},
\label{eq:G(I)}
\end{equation}
where the second term in the brackets explicitly removes the tilt of the distribution and we have defined $\Delta\mathcal{I}_\pm=\mathcal{I}-\langle I_\pm\rangle$. From Eq.~(\ref{eq:P(I)}), we then obtain the semi-circle
\begin{equation}
\mathcal{G}^2(\mathcal{I}) + \left(\frac{\mathcal{I}-\bar I}{\langle I_+\rangle-\langle I_-\rangle}\right)^2=1,
\label{eq:G(I)_circle}
\end{equation}
which should describe the fluctuations in any bistable conductor independently of the microscopic details. Here we have defined $\bar{I}=(\langle I_+\rangle+\langle I_-\rangle)/2$. Figure~\ref{fig:3}b shows that our experimental data in Fig.~\ref{fig:3}a measured at long times indeed collapse onto this semi-circle when rescaled according to Eq.~(\ref{eq:G(I)}). This property should hold for a variety of bistable systems from different fields of physics.

\emph{Fluctuation relation.---} Finally, we examine the symmetry properties of the fluctuations \cite{Esposito2009}. Equation~(\ref{eq:P(I)}) is suggestive of a fluctuation relation at long times reading
\begin{equation}
\frac{1}{\tau}\ln \left[\frac{P_\tau(\mathcal{I}=\bar{I}+\mathcal{J})}{P_\tau(\mathcal{I}=\bar{I}-\mathcal{J})}\right]=\Omega\mathcal{J},
\label{eq:FT}
\end{equation}
where $\Omega =2(\Gamma_+ - \Gamma_-)/(\langle I_+\rangle-\langle I_-\rangle) $ controls the slope. This relation is reminiscent of the Gallavotti-Cohen fluctuation theorem \cite{Gallavotti1995,Gallavotti1995a}, however, here the intensive entropy production is replaced by the departure $\mathcal{J}$ from the average $\bar{I}$ of the mean currents. Equation~(\ref{eq:FT}) should be valid in the bistable region of the distribution which is dominated by the random tunneling in the SEB. The excellent agreement between theory and experiment in Fig.~\ref{fig:4} confirms the prediction. The fluctuation relation is expected to be valid for many different bistable systems and may be further tested in future experiments.

\emph{Conclusions.---} We have realized a controllable bistability in order to investigate fundamental properties of current fluctuations in bistable conductors. These include the cross-over from short-time to long-time statistics, the large-deviation principle, and the fluctuation relation in long-time limit. Our results have an immediate application for the detection and characterization of spurious two-level fluctuators which appear in many solid-state devices, since their fluctuations are universal and independent of the microscopic details. We have formulated and verified universal properties including a fluctuation relation for bistable conductors. Our work establishes several analogies between bistable conductors and concepts from statistical mechanics and it offers perspectives for further experiments on statistical physics with electronic conductors.

\emph{Acknowledgments.---} We thank Yu.~Galperin, F.~Ritort, and K.~Saito for useful discussions. The work was supported by Academy of Finland (projects 273827, 275167, and 284594) and Foundation Nanosciences under the aegis of Joseph Fourier University Foundation. We acknowledge the provision of facilities by Aalto University at OtaNano -- Micronova Nanofabrication Centre. Authors at Aalto University are affiliated with Centre for Quantum Engineering.



\section*{Supplementary material}

# Distribution of Current Fluctuations in a Bistable Conductor
## Supplemental material

S. Singh,[1] J. T. Peltonen,[1] I. M. Khaymovich,[1,2,3] J. V. Koski,[1,4] C. Flindt,[1] and J. P. Pekola[1]

[1] Low Temperature Laboratory, Department of Applied Physics, Aalto University, 00076 Aalto, Finland
[2] Institute for Physics of Microstructures, Russian Academy of Sciences, 603950 Nizhny Novgorod, GSP-105, Russia
[3] LPMMC, Université de Grenoble-Alpes and CNRS, 38042 Grenoble, France
[4] Solid State Physics Laboratory, ETH Zurich, 8093 Zurich, Switzerland

## I. MASTER EQUATION

We consider a two-state system in the framework of a master equation for the probabilities $p_n(t)$ to be in the charge state $n = 0$ or $n = 1$ at time $t$. We focus on the probability distribution $\mathcal{P}(\mathcal{N})$ for the time-averaged value of $n$,

$$\mathcal{N}(\tau) = \frac{1}{\tau} \int_0^\tau n(t) dt \tag{1}$$

using a large-deviation approach (see Ref. 1 for review). Thus, we go beyond the central-limit theorem and consider the rare events in the tails of the distribution together with the scaling of $\mathcal{P}(\mathcal{N})$ with $\tau$. The master equation reads

$$\dot{p}_1(t) = \Gamma_+ - (\Gamma_+ + \Gamma_-)p_1(t), \tag{2}$$

where $\Gamma_\pm$ are the tunneling rates to ($n = 0 \to 1$) and from the island ($n = 1 \to 0$), respectively. The probabilities are normalized such that $p_0(t) + p_1(t) = 1$ at all times.

## II. MULTI-JUMP APPROACH

The probability of realizing a trajectory, in the time interval from $t_0 = 0$ to $t_{M+1} = \tau$ with $M$ alternating jumps of the form $n = 0 \to 1$ and $n = 1 \to 0$ at times $t_1 < \ldots < t_M$, reads[2]

$$\mathcal{P}_M^\pm(t_1, \ldots, t_M) = p_{0,1}(0) \exp\left[ -\sum_{j=0}^M \Gamma_{\pm(-1)^j}(t_{j+1} - t_j) \right] \prod_{j=0}^{M-1} \Gamma_{\pm(-1)^j} . \tag{3}$$

Here, the initial state $n(0) = 0$ (1) corresponds to the upper (lower) signs. In the stationary state, the initial probabilities $p_n(0)$ are

$$p_0(0) = \frac{\Gamma_-}{\Gamma_{\text{tot}}} \quad \text{and} \quad p_1(0) = \frac{\Gamma_+}{\Gamma_{\text{tot}}} , \tag{4}$$

where $\Gamma_{\text{tot}} = \Gamma_+ + \Gamma_-$ is the total relaxation rate. Based on Eq. (3), one can write the time-average in Eq. (1) as

$$\mathcal{N} = \frac{1}{\tau} \sum_{j=1}^M (-1)^{j+\eta} t_j + \text{odd}(M + \eta), \tag{5}$$

where $n(0) = \eta$ is given by initial state, and $\text{odd}(m) = 0$ for even $m$ and $\text{odd}(m) = 1$ for odd $m$.

The probability distribution of $\mathcal{N}$ can now be expressed with the help of Eq. (3) as

$$\mathcal{P}(\mathcal{N}) = \sum_{M=0}^\infty \sum_{\eta=\pm} \int \cdots \int_{0 < t_1 < \ldots < t_M < \tau} \delta\left( \mathcal{N} - \frac{1}{\tau} \sum_{j=1}^M (-1)^{j+\eta} t_j + \text{odd}(M + \eta) \right) \mathcal{P}_M^\eta(t_1, \ldots, t_M) \prod_{j=1}^M dt_j, \tag{6}$$

where the $M$-jump probabilities are written as

$$\mathcal{P}_M^\pm(t_1, \ldots, t_M) = \frac{1}{\tau^M} p_{0,1}(0) e^{\gamma_d \mathcal{N} - \gamma_+} (\gamma_+ \gamma_-)^{[M/2]} (\gamma_\pm)^{\text{odd}(M)} . \tag{7}$$

Here we have introduced the dimensionless tunneling rates $\gamma_j = \Gamma_j \tau$ and $\Gamma_d = \Gamma_+ - \Gamma_-$, and $[x]$ denotes the integer part of $x$. To proceed, we need to evaluate the following integrals

$$J_M^\eta(\mathcal{N}) = \int \cdots \int\limits_{0 < t_1 < \ldots < t_M < \tau} \delta \left( \mathcal{N} - \frac{1}{\tau} \sum_{j=1}^M (-1)^{j+\eta} t_j + \text{odd}(M + \eta) \right) \prod_{j=1}^M dt_j \ . \tag{8}$$

To this end we change variables as follows

$$D_1 = t_1, \quad D_m = t_m - D_{m-1} = \sum_{j=1}^m t_j (-1)^{M-j} \ . \tag{9}$$

The inequalities $0 < t_1 < \ldots < t_M < \tau$ lead to a splitting of the set $\{D_m\}$ into two ascending sequences (one for even and one for odd $m$'s),

$$D_{m-1} < D_{m+1} \tag{10}$$

with the following conditions on the first and last terms in the sequences

$$D_1, D_2 > 0 \quad \text{and} \quad D_M + D_{M-1} < \tau. \tag{11}$$

According to Eq. (9), which can be written as $t_m = D_m + D_{m-1}$, the Jacobian $\partial(t_1, \ldots, t_M)/\partial(D_1, \ldots, D_M)$ is unity.

The expression for $\mathcal{P}(\mathcal{N})$ can be represented in terms of the $D_M$'s, since

$$\frac{1}{\tau} \sum_{j=1}^M (-1)^{j+\eta} t_j + \text{odd}(M + \eta) = \frac{1}{\tau} (-1)^{M+\eta} D_M + \text{odd}(M + \eta) = \frac{1}{\tau} \left\{ \begin{array}{ll} D_M, & M = 2k \\ \tau - D_M, & M = 2k+1 \end{array} \right. . \tag{12}$$

Moreover, time-reversal symmetry ($t_m \to \tau - t_{M-m+1}$) for $M = 2k+1$ together with the symmetry ($n \to 1 - n$) imply for the $J_M^\eta$ that

$$J_{2k+1}^0(\mathcal{N}) = J_{2k+1}^1(\mathcal{N}) \quad \text{and} \quad J_M^0(\mathcal{N}) = J_M^1(1 - \mathcal{N}). \tag{13}$$

It then suffices to evaluate $J_M^{(\eta=0)}$ for which we find

$$J_{2k}^0(\mathcal{N}) = \int \cdots \int\limits_{0 < D_2 < D_4 \ldots < D_{2k} < \tau} \delta \left( \mathcal{N} - \frac{1}{\tau} D_{2k} \right) \prod_{j=1}^k dD_{2j} \int \cdots \int\limits_{0 < D_1 < D_3 \ldots < D_{2k-1} < \tau - D_{2k}} \prod_{j=1}^k dD_{2j-1} = \tau^M \frac{\mathcal{N}^{k-1}}{(k-1)!} \frac{(1 - \mathcal{N})^k}{k!} \tag{14}$$

and

$$J_{2k+1}^0(\mathcal{N}) = \int \cdots \int\limits_{0 < D_1 < D_3 \ldots < D_{2k+1} < \tau} \delta \left( \mathcal{N} - 1 + \frac{1}{\tau} D_{2k+1} \right) \prod_{j=0}^k dD_{2j+1} \int \cdots \int\limits_{0 < D_2 < D_4 \ldots < D_{2k} < \tau - D_{2k+1}} \prod_{j=1}^k dD_{2j} = \tau^M \frac{(1 - \mathcal{N})^k}{k!} \frac{\mathcal{N}^k}{k!}. \tag{15}$$

As a result we obtain

$$\begin{aligned}
\mathcal{P}(\mathcal{N}) &= e^{\gamma_d \mathcal{N} - \gamma_+} \left\{ p_0(0)\delta(\mathcal{N}) + p_1(0)\delta(\mathcal{N} - 1) + \sum_{k=0}^\infty (\gamma_+ \gamma_-)^k \frac{\mathcal{N}^k}{k!} \frac{(1 - \mathcal{N})^k}{k!} \sum_{\eta=0}^1 \left[ p_0(0)\gamma_+^\eta \left( \frac{k}{\mathcal{N}} \right)^{1-\eta} + p_1(0)\gamma_-^\eta \left( \frac{k}{1 - \mathcal{N}} \right)^{1-\eta} \right] \right\} \\
&= \frac{e^{\gamma_d \mathcal{N} - \gamma_+}}{\gamma_{\text{tot}}} \left\{ \gamma_-\delta(\mathcal{N}) + \gamma_+\delta(\mathcal{N} - 1) + \sum_{k=0}^\infty (\gamma_+ \gamma_-)^k \frac{\mathcal{N}^k}{k!} \frac{(1 - \mathcal{N})^k}{k!} \left[ 2\gamma_+ \gamma_- + k \frac{\gamma_- + \gamma_d \mathcal{N}}{\mathcal{N}(1 - \mathcal{N})} \right] \right\} \\
&= \frac{e^{\gamma_d \mathcal{N} - \gamma_+}}{\gamma_{\text{tot}}} \left\{ \gamma_-\delta(\mathcal{N}) + \gamma_+\delta(\mathcal{N} - 1) + \sum_{k=0}^\infty (\gamma_+ \gamma_-)^{k+1} \frac{\mathcal{N}^k (1 - \mathcal{N})^k}{k!(k+1)!} \left[ 2(k+1) + (\gamma_- + \gamma_d \mathcal{N}) \right] \right\}
\end{aligned} \tag{16}$$

with $M = 2k + \eta$. In the last step, we have skipped a term which is zero for $M = 0$ and have put $M = 2k + 1$ and $M = 2k + 2$ together into the square brackets. The sums in the last expression are simply the modified Bessel functions $I_p(x)$ with the common argument $x = 2\sqrt{\gamma_+ \gamma_- \mathcal{N}(1 - \mathcal{N})}$. We then arrive at the expression

$$\mathcal{P}(\mathcal{N}) = \frac{e^{\gamma_d \mathcal{N} - \gamma_+}}{\gamma_{\text{tot}}} \left\{ \gamma_-\delta(\mathcal{N}) + \gamma_+\delta(\mathcal{N} - 1) + 2\gamma_+ \gamma_- \theta(\mathcal{N})\theta(1 - \mathcal{N}) \left[ I_0(x) + (\gamma_- + \gamma_d \mathcal{N}) \frac{I_1(x)}{x} \right] \right\}, \tag{17}$$

which is Eq. (3) of the main text.

### III. LIMITING CASES

In the limit of short integration times $\tau$ (implying that $\gamma_k \to 0$), only the first two terms corresponding to zero-jump trajectories survive in the expression for the probability distribution $\mathcal{P}(\mathcal{N})$

$$\mathcal{P}(\mathcal{N})|_{\tau \to 0} = \frac{\gamma_-}{\gamma_{\text{tot}}} \delta(\mathcal{N}) + \frac{\gamma_+}{\gamma_{\text{tot}}} \delta(\mathcal{N} - 1). \tag{18}$$

In the opposite limit of large $\tau$ ($\gamma_k \gg 1$), we find

$$\mathcal{P}(\mathcal{N}) \simeq \frac{e^{\gamma_d \mathcal{N} - \gamma_+}}{\gamma_{\text{tot}}} 2\gamma_+ \gamma_- \left[ 1 + \frac{\gamma_- + \gamma_d \mathcal{N}}{2\sqrt{\gamma_+ \gamma_- \mathcal{N}(1 - \mathcal{N})}} \right] \frac{e^{2\sqrt{\gamma_+ \gamma_- \mathcal{N}(1 - \mathcal{N})}}}{(4\pi)^{1/2} (\gamma_+ \gamma_- \mathcal{N}(1 - \mathcal{N}))^{1/4}} \tag{19}$$

or in the logarithmic form

$$\frac{\ln \mathcal{P}(\mathcal{N}) - \gamma_d \mathcal{N} + \gamma_+}{\sqrt{\gamma_+ \gamma_-}} \simeq 2\sqrt{\mathcal{N}(1 - \mathcal{N})} + \frac{1}{\sqrt{\gamma_+ \gamma_-}} \ln \left[ \frac{2\gamma_+ \gamma_-}{\gamma_{\text{tot}} (4\pi)^{1/2} (\gamma_+ \gamma_- \mathcal{N}(1 - \mathcal{N}))^{1/4}} \left( 1 + \frac{\gamma_- + \gamma_d \mathcal{N}}{2\sqrt{\gamma_+ \gamma_- \mathcal{N}(1 - \mathcal{N})}} \right) \right]. \tag{20}$$

Here we have used the asymptotic behavior of the Bessel functions $I_0(x) \simeq I_1(x) \simeq e^x / \sqrt{2\pi x}[1 + O(1/x)]$ for large $x$.

### IV. ADDITION OF DELTA-CORRELATED NOISE

In the experiment, the state $n(t)$ is measured together with the noise $\zeta(t)$ at the time instants $t_k = k \cdot \Delta t$, i. e.

$$\tilde{n}(t_k) = n(t_k) + \zeta_k. \tag{21}$$

The noise $\zeta_k = \zeta(t_k)$ is stationary and delta-correlated with the distribution function $P_0(\zeta)$. To calculate the distribution $\widetilde{\mathcal{P}}(\widetilde{\mathcal{N}})$ of the variable $\widetilde{\mathcal{N}}$ given by Eq. (1) with $n(t)$ substituted by $\tilde{n}(t)$, we need the convolution of the distribution function $\mathcal{P}(N)$ in Eq. (17) and the distribution function $\mathcal{P}_\xi(\xi(\tau))$ of the averaged noise

$$\xi(\tau) = \frac{1}{\tau} \int \zeta(t) dt = \frac{1}{K} \sum_{k=1}^{K} \zeta_k, \tag{22}$$

where $K = \tau / \Delta t$. This convolution reads

$$\widetilde{\mathcal{P}}(\widetilde{\mathcal{N}}) = \int_0^1 \mathcal{P}(\mathcal{N}) d\mathcal{N} \int \mathcal{P}_\xi(\xi(\tau)) d\xi(\tau) \delta(\widetilde{\mathcal{N}}) - \mathcal{N} - \xi(\tau)) = \int_0^1 \mathcal{P}(\mathcal{N}) \mathcal{P}_\xi(\widetilde{\mathcal{N}} - \mathcal{N}) d\mathcal{N} =$$

$$= \frac{\gamma_- e^{-\gamma_+} \mathcal{P}_\xi(\widetilde{\mathcal{N}}) + \gamma_+ e^{-\gamma_-} \mathcal{P}_\xi(\widetilde{\mathcal{N}} - 1)}{\gamma_{\text{tot}}} + \frac{2\gamma_+ \gamma_-}{\gamma_{\text{tot}}} \int_0^1 e^{\gamma_d \mathcal{N} - \gamma_+} \left[ I_0(x) + (\gamma_- + \gamma_d \mathcal{N}) \frac{I_1(x)}{x} \right] \mathcal{P}_\xi(\widetilde{\mathcal{N}} - \mathcal{N}) d\mathcal{N}. \tag{23}$$

In addition, we have the following expression for the generating function of the variable $K \cdot \xi(\tau)$

$$\langle e^{iK\xi(\tau)q} \rangle = \int e^{iK\xi(\tau)q} \mathcal{P}_\xi(\xi(\tau)) d\xi(\tau) = \prod_{k=1}^{K} \int e^{i\xi_k q} P_0(\xi_k) d\xi_k = \left[ \int e^{i\zeta q} P_0(\zeta) d\zeta \right]^K. \tag{24}$$

Here, we have used that the noise is delta-correlated ($\zeta_k$ and $\zeta_l$ are independent for $k \neq l$) and stationary (the distribution $P_0(\zeta_k)$ is the same for all $k$). Finally, by an inverse Fourier transformation we obtain $\mathcal{P}_\xi(\xi(\tau))$ on the form

$$\mathcal{P}_\xi(\xi(\tau)) = \frac{1}{2\pi} \int e^{-iK\xi(\tau)q} \langle e^{iK\xi(\tau)q} \rangle K dq = \frac{K}{2\pi} \int e^{-iK\xi(\tau)q} \left[ \int e^{i\zeta q} P_0(\zeta) d\zeta \right]^K dq. \tag{25}$$

### A. Gaussian and Lorentzian distributions

We consider the noise $\xi$ to be either Gaussian or Lorentzian distributed. For the Gaussian distribution, we have

$$P_0(\zeta) = \frac{1}{\sqrt{2\pi}\sigma}e^{-\zeta^2/2\sigma^2}.$$  (26)

We then find

$$\int e^{i\zeta q}P_0(\zeta)d\zeta = \frac{1}{\sqrt{2\pi}\sigma}\int e^{-\zeta^2/2\sigma^2+i\zeta q}d\zeta = e^{-\sigma q^2}$$  (27)

and

$$\mathcal{P}_\xi(\xi(\tau)) = \frac{K}{2\pi}\int e^{-iK(\xi(\tau)q-\sigma q^2)}dq = \frac{1}{\sqrt{2\pi/K}\sigma}e^{-\xi(\tau)^2/2(\sigma^2/K)}$$  (28)

which is also a Gaussian with the variance $\sigma^2/K$.

For the Lorenzian distribution

$$P_0(\zeta) = \frac{1}{\pi}\frac{\sigma}{\zeta^2+\sigma^2},$$  (29)

we obtain

$$\int e^{i\zeta q}P_0(\zeta)d\zeta = \frac{\sigma}{\pi}\int e^{i\zeta q}\frac{d\zeta}{\zeta^2+\sigma^2} = e^{-\sigma|q|}$$  (30)

and

$$\mathcal{P}_\xi(\xi(\tau)) = \frac{K}{2\pi}\int_{-\infty}^{\infty}e^{-K(i\xi(\tau)q+\sigma|q|)}dq = \frac{K}{\pi}\text{Re}\int_0^\infty e^{-Kq(i\xi(\tau)+\sigma)}dq = \frac{1}{\pi}\frac{\sigma}{\xi(\tau)^2+\sigma^2}$$  (31)

which coincides with the initial noise distribution. In both cases, one should substitute Eq. (28) or Eq. (31) into Eq. (23) and then evaluate the resulting convolution.

---



\end{document}